\documentclass[twocolumn,aps,prapplied,longbibliography,showpacs,superscriptaddress]{revtex4-2}
\usepackage{amsfonts}
\usepackage{amsmath}
\usepackage{amssymb}
\usepackage{graphicx}
\usepackage{dcolumn}
\usepackage{bm}
\usepackage[utf8]{inputenc}
\usepackage[T1]{fontenc}
\usepackage{mathptmx}
\usepackage{etoolbox}
\usepackage{color}

\newcommand{\mycol}{1}

\draft 

\begin{document}
	
	\preprint{}
	
	\title{Strong coupling of a Gd$^{3+}$ multilevel spin system to an on-chip superconducting resonator} 
	
	
	
	\author{Giovanni~Franco-Rivera}
	\email{gf15@fsu.edu}
	\affiliation{Department of Physics and The National High Magnetic Field Laboratory, Florida State University, Tallahassee, Florida 32310, USA}
	
	\author{Josiah Cochran}
	\affiliation{Department of Physics and The National High Magnetic Field Laboratory, Florida State University, Tallahassee, Florida 32310, USA}
	
	\author{Seiji~Miyashita}
	\affiliation{Department of Physics, Graduate School of Science, The University of Tokyo, 7-3-1 Bunkyo-Ku, Tokyo, 113-0033 Japan}
	
	\author{Sylvain~Bertaina}
	\affiliation{CNRS, Aix-Marseille Universit\'{e}, IM2NP (UMR 7344), Institut Mat\'{e}riaux Micro\'{e}lectronique et Nanosciences de Provence, Marseille, France}
	
	\author{Irinel~Chiorescu}
	\email{ichiorescu@fsu.edu}
	\affiliation{Department of Physics and The National High Magnetic Field Laboratory, Florida State University, Tallahassee, Florida 32310, USA}
	
	\date{\today}
	
	\begin{abstract}
		We report the realization of a strong coupling between a Gd$^{3+}$ spin ensemble hosted in a scheelite (CaWO$_4$) single crystal and the resonant mode of a coplanar stripline superconducting cavity leading to a large separation of spin-photon states of 146~MHz. The interaction is well described by the Dicke model and the crystal-field Hamiltonian of the multilevel spin system. We observe a change of the crystal-field parameters due to the presence of photons in the cavity that generates a significant perturbation of the crystal ground state. Using finite-element calculations, we numerically estimate the cavity sensing volume as well as the average spin-photon coupling strength of $g_0\approx$~620~Hz. Lastly, the dynamics of the spin-cavity states are explored via pulsed measurements by recording the cavity ring-down signal as a function of pulse length and amplitude. The results indicate a potential method to initialize this multilevel system in its ground state via an active cooling process. 
	\end{abstract}
	
	
	\maketitle 
	
	
	
	%
	%
\section{Introduction}	
Interaction between quantum systems via electromagnetic excitations are currently the basis of operation of many quantum hybrid systems using photonic entanglement~\cite{Kurizki2015}. The confinement of electromagnetic fields in mesoscopic mode volumes, as in the case of on-chip superconducting cavities, allows the study of light and matter interactions in the strong-coupling regime when its coupling strength $g_c$ exceeds the qubit dephasing rate $\gamma$ and cavity dissipation rate $\kappa_c$ \cite{Haroche2006}. Experimentally, electric dipole coupling between electromagnetic excitations in a cavity and a single quantum emitter has been achieved in atomic systems \cite{Thompson1992} and superconducting qubits~\cite{Wallraff2004} where the electric dipole interaction with the cavity mode is large. In contrast, achieving strong magnetic coupling between an electromagnetic field and a single quantum spin remains elusive due to its small magnetic dipole. However, the interaction can be collectively enhanced by employing an ensemble of $N$-identical spins such that the ensemble coupling strength is enhanced by a factor of $\sqrt{N}$. In this way strong coupling has been demonstrated with various spin systems like N-V centers and point defects in diamond~\cite{Kubo2010a, Schuster2010a, Amsuss2011}, molecular magnets \cite{Chiorescu2010, Eddins2014, Bonizzoni2018} and transition metals and rare-earth (RE) ions in crystals~\cite{Schuster2010a,Bushev2011, Probst2013, Keyser2020, Dold2019,Wang2022}. Moreover, recent studies using Er$^{3+}$:Y$_2$SiO$_5$ spin diluted crystals~\cite{Probst2015prb}, N-V centers in diamond~\cite{Grezes2015} and Bi defects in Si~\cite{Ranjan2020prl} demonstrated storing and retrieving the state of microwave photons at high-power and near the quantum limit regime.

Among the 4f ions, Gd$^{3+}$ is half-filled and posses the largest spin quantum number ($S=7/2$) with no orbital angular momentum ($L=0$). When doped into the CaWO$_4$, this spin provides a large magnetic moment and crystal field which in moderate fields ($\sim0.1$~kG) allows spin transitions at large frequencies of $\approx$18~GHz. The characteristics of the crystal field allow the use of clock transitions where spin coherence is insensitive to field fluctuations~\cite{Franco-Rivera2022}. It is important to note that the higher the transition (or cavity) frequency, the higher the spin-photon coupling which is another benefit of the Gd$^{+3}$ crystal field. Nevertheless, the resulting spin dynamics is using a frequency regime still practical for integration in circuit Quantum Electro-Dynamics (QED) architectures.

By performing field dependent cavity spectroscopy measurements we explore the coupling between a Gd$^{3+}$ spin ensemble to a superconducting on-chip resonator. We demonstrate the strong-coupling regime characterized by a large avoided crossing of 146 MHz, so large that it leads to a significant perturbation of crystal's ground state by the presence of a photon. The experimentally observed spin-cavity dressed states are well described by the Dicke model \cite{Dicke1954} for a multi-level system with high anisotropy. Moreover, we perform pulsed electron spin resonance (ESR) measurements which show a population inversion and suggest a way to perform active cooling of the spin system into its ground state.

\section{Observation of the strong coupling regime described by the Dicke model}
\begin{figure}
	\centering
	\includegraphics[width=\mycol\columnwidth]{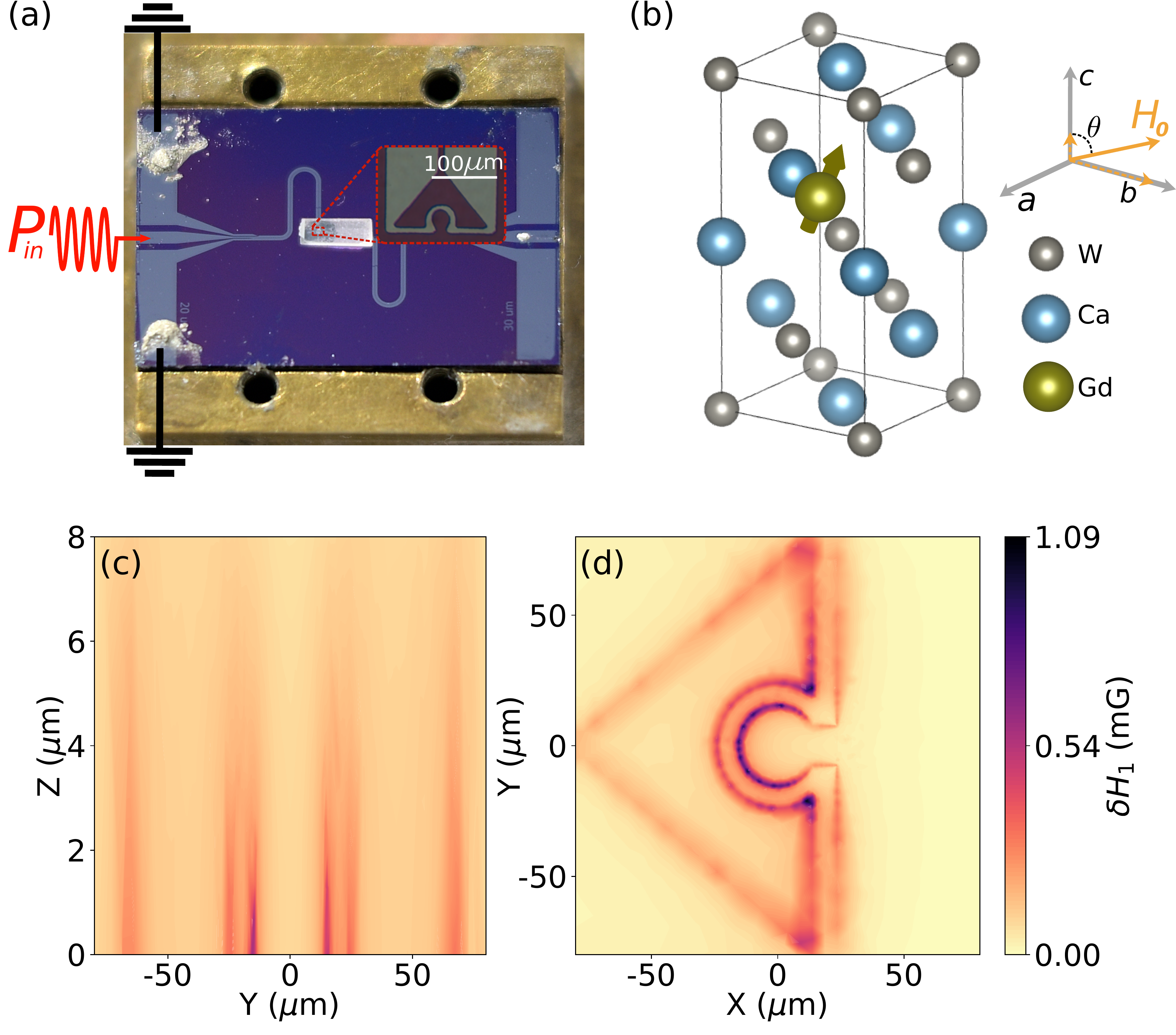}
	\caption{(a) Coplanar stripline resonator photograph with Gd$^{3+}$ spin sample. Inset: optical micrograph of the $\Omega$-shaped short circuit. (b) Unit cell of the Gd$^{3+}$ doped CaWO$_4$ crystal; the O atoms are removed for clarity. A static field \textit{H}$_0$ is applied at an angle $\theta$ = 81.46\textdegree~relative to the crystallographic \textit{c}~axis. (c-d) COMSOL simulations of the $\delta H_1^{rms}$-field distribution (vacuum field corresponding to $P_{rms}=-124.1$~dBm) at (c) the center of the $\Omega$-shaped loop vs distance $z$ from the chip and (d) at the chip surface $z=0$ showing where $\delta H_1^{rms}$-field is mostly concentrated.} 
	\label{fig1}
\end{figure}
The spectroscopy of resonator-photon spin interaction is studied using a home-built heterodyne detector~\cite{Franco-Rivera2022}. The sample holder is introduced inside a superconducting vector magnet with field $\boldsymbol{H}_0$ and thermally anchored to the mixing chamber of a dilution refrigerator at $T\simeq0.38$~K. The resonator is etched out of a 20 nm Nb film and contains a transition from a coplanar waveguide (CPW) to a coplanar stripline (CPS) \cite{Anagnostou2008} matched for a 50~$\Omega$ impedance. The CPS is capacitively coupled to a $\lambda/4$ resonant structure ending with a short-circuit shaped into an $\Omega$-loop (internal radius of 15~$\mu$m). Fig.~\ref{fig1}a shows a single crystal (size $\simeq 2.5\times1\times 0.6$~mm$^3$) of CaWO$_4$ with a 0.05\% spin concentration of Gd$^{3+}$ (Fig.~\ref{fig1}b) well pressed on a bead of grease atop of the loop, where the field distribution of the resonant mode is mostly concentrated (see Fig.~\ref{fig1}c-d). The crystal has a $I4/a$ tetragonal symmetry with lattice constants of the unit cell $a=b=$~5.24~\AA~and $c=$~11.38~\AA.

The electron spin Hamiltonian of the Gd$^{3+}$ $S=7/2$ ion is given by~\cite{Hempstead1960, Zalkin1964}:
\begin{equation}
	\mathcal{H}_s = \mu_B \boldsymbol{H}^{T}_0 \boldsymbol{g}\boldsymbol{S} + B_2^0 O_2^0 + B_4^0 O_4^0 + B_4^4 O_4^4 + B_6^0 O_6^0 + B_6^4 O_6^4
	\label{gdhamiltonian}
\end{equation}
where $O_k^q$ are the Stevens' operators~\cite{Rudowicz2004}, $\mu_B$ is the Bohr magneton, $\boldsymbol{H}_0$ is the applied magnetic field vector and ($g_{\|}$, $g_{\perp}$) = (1.991,1.992) are the g-factors parallel and perpendicular to the crystallographic \textit{c}-axis respectively.

\begin{figure}
	\centering
	\includegraphics[width=\mycol\columnwidth]{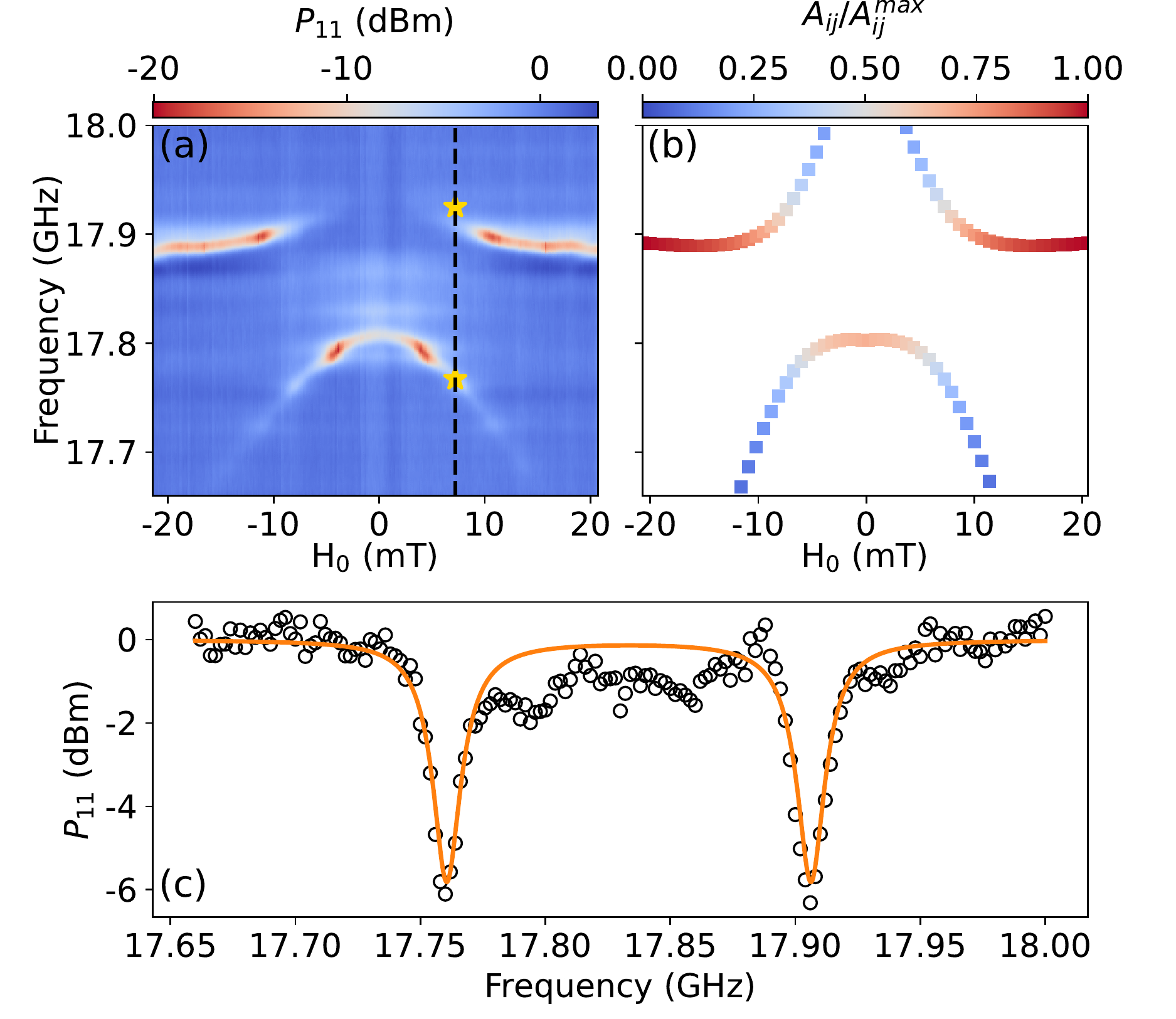}
	\caption{(a) Reflected power as a function of field and frequency showing a large splitting of $\approx$~146~MHz due to spin-photon strong coupling. The dashed line indicates the location of maximum separation between dips. (b) Normalized $A_{2,4}$ and $A_{2,5}$ (see Eq.~\ref{eq:Aif}) as the lower and upper branches, respectively. Simulated resonance frequencies are shown by yellow stars on panel (a), in good agreement with the experiment. (c) Reflected power $P_{11}$ (black circles) at the resonance field represented by the dashed line in panel (a). The solid orange line corresponds to a fit to the data using Eq.~\ref{eq:s11spin}.} 
	\label{fig2}
\end{figure}

Measurements of the field and frequency dependent reflected microwave power are given in Fig. \ref{fig2}a. The spin-resonator system is probed with a long 10 $\mu$s pulse at a 5~ms repetition time, to allow spin relaxation before each pulse. The repetition time is on par with relaxation time measurements done at 6~K \cite{Baibekov2017}. A sequence of 320 pulses are sent and the averaged Fast Fourier Transform (FFT) of the reflected signal ringdown is recorded. At the sample holder level we estimate the power to be $P_{in}=-64$~dBm (see Fig.\ref{fig1}a). For a fixed field $H_0$ the excitation frequency $\omega$ is swept around the cavity resonance and the process is repeated from negative to positive $H_0$ values. A small background signal is recorded away from the spin-cavity resonance and subtracted from the reflected power, in order to remove unwanted resonances attributed to impedance mismatch between the coaxial line and sample holder. A strong hybridization between the spin and cavity modes results in the observed avoided crossing  at $H_r\approx \pm$0.072~kG where the splitting has the largest value (146~$\pm$~1~MHz). As described below, we used an exact diagonalization and least square algorithm to fit the data and the simulated spectrum is in good agreement with the experimental results (see Fig.~\ref{fig2}b).

Figure~\ref{fig2}c shows reflected power $P_{11}(\frac{\omega}{2\pi})$ (in dBm) for $\omega_s = \omega_c$ where the two reflection dips have equal intensity and largest split ($\omega_s$ is the spin levels separation and $\omega_c$ is the resonator mode). The fit (orange line) of the reflection scattering coefficient is based on input-output formalism in the presence of a spin system~\cite{Abe2011, Schuster2010a} ($P_{11}\propto S_{11}^2$):
\begin{equation}
	\left|S_{11}\right|^2 (\omega)= \left|1 + \frac{\kappa_e}{i\left(\omega - \omega_c\right) - \kappa_c + \frac{g_c^2}{i\left(\omega-\omega_s\right)-\gamma_s}}\right|^2. 		
	\label{eq:s11spin}
\end{equation}
The cavity $\kappa_c/(2\pi)$ = 10.485~MHz and external $\kappa_e/(2\pi)\approx$~0.99$\kappa_c/(2\pi)$ = 10.38~MHz damping rates of the resonator are determined from the reflected spectrum away from resonance and used as fit constants. We obtain an ensemble spin-resonator coupling strength $g_c/(2\pi)=$~73.0$\pm$0.6~MHz and a spin linewidth (dephasing rate) $\gamma_s/(2\pi)=$~8.8$\pm$1.5~MHz. The $g_c$ value is simply half of the signal splitting and is an oversimplified view when translated to the Gd multi-level system. The coupling term between the Gd$^{3+}$ spin $S=7/2$ and the photon mode has to be described by the Dicke model~\cite{Dicke1954} initially proposed to describe the superradiance emission of light by an ensemble of $N$ atoms. We apply it here for a spin $S = N/2$, with $N=7$ in the presence of the CaWO$_4$ crystal field:
 \begin{equation}
 	\mathcal{H} = \mathcal{H}_{s} + \frac{\omega_c}{2\pi} a^{\dagger}a + \frac{g_c}{2\pi}\left(a^{\dagger} + a\right)\left(S^{+} + S^{-}\right),
 	\label{eq:gdcavH}
 \end{equation}
where $a^\dagger, a$ are the photon creation and annihilation operators and $S_{\pm}$ are the spin raising/lowering operators, respectively.
\begin{table*}[t]
	\renewcommand{\arraystretch}{1.2}
	\footnotesize
	\centering
	\caption{Eigenstates of the cavity-spin Hamiltonian at the observed resonance field $H_r=72$~G. Amplitude coefficients less than 0.01 have been disregarded for clarity.}
	\begin{tabular*}{2\columnwidth}{ccccccccc}
		\hline\hline
		\\
		\centering
		\hspace{1mm}$|3\rangle = $ \hspace{5mm} & \hspace{1mm} $S_z= 7/2$ \hspace{5mm} & \hspace{1mm} $S_z = 5/2$ \hspace{5mm} & \hspace{1mm} $S_z = 3/2$ \hspace{5mm} & \hspace{1mm} $S_z = 1/2$ \hspace{5mm} & \hspace{1mm} $S_z = -1/2$ \hspace{5mm} & \hspace{1mm} $S_z = -3/2$ \hspace{5mm} & \hspace{1mm} $S_z -5/2$ \hspace{5mm} & \hspace{1mm} $S_z = -7/2$ \hspace{5mm} \\
		\hline
		$n = 0$ & 0.010 & -0.629 & -0.092 & 0 & 0 & -0.153 & -0.318 & 0 \\
		$n = 1$ & 0.669 & 0.01 & 0 & 0 & 0.036 & 0 & 0 & 0.145 \\
		\hline
		\\
		\hspace{1mm}$|4\rangle = $ \hspace{5mm} & \hspace{1mm} $S_z= 7/2$ \hspace{5mm} & \hspace{1mm} $S_z = 5/2$ \hspace{5mm} & \hspace{1mm} $S_z = 3/2$ \hspace{5mm} & \hspace{1mm} $S_z = 1/2$ \hspace{5mm} & \hspace{1mm} $S_z = -1/2$ \hspace{5mm} & \hspace{1mm} $S_z = -3/2$ \hspace{5mm} & \hspace{1mm} $S_z -5/2$ \hspace{5mm} & \hspace{1mm} $S_z = -7/2$ \hspace{5mm} \\
		\hline
		$n = 0$ & 0 & 0.145 & 0.140 & 0 & 0 & 0.05 & 0.578 & 0 \\
		$n = 1$ & 0.561 & 0 & 0 & -0.03 & 0.03 & 0 & 0 & -0.553 \\
		\hline
		\\
		
		\hspace{1mm}$|5\rangle = $ \hspace{5mm} &\hspace{1mm} $S_z = 7/2$ \hspace{5mm} & \hspace{1mm} $S_z = 5/2$ \hspace{5mm} & \hspace{1mm} $S_z = 3/2$ \hspace{5mm} & \hspace{1mm} $S_z = 1/2$ \hspace{5mm} & \hspace{1mm} $S_z = -1/2$ \hspace{5mm} & \hspace{1mm} $S_z = -3/2$ \hspace{5mm} & \hspace{1mm} $S_z=-5/2$ \hspace{5mm} & \hspace{1mm} $S_z=-7/2$ \hspace{5mm} \\
		\hline
		$n = 0$ & 0 & 0.605 & 0.015 & 0 & 0 & 0.141 & -0.010 & 0 \\
		$n = 1$ & 0.460 & 0 & 0 & 0.033 & 0.024 & 0 & 0 & 0.631 \\
		
		\hline
	\end{tabular*}
	\label{table1}
\end{table*}

Contrary to the Tavis-Cumming model the inclusion of the counter-rotating coupling terms is necessary for a spin $S=7/2$ in the presence of a crystal field where the zero-field splitting energy is comparable to the resonator transition frequency. As discussed below, the number of photons in the cavity $n_c = a^{\dagger}a$ is much smaller than the number of excited spins and therefore we can use an exact diagonalization of $\mathcal{H}$ for $n_c=0,1$, which is a $2(2S+1)$ matrix. The spin-photon eigenvalues/states $\{E_k,|k\rangle\}_{k=1\ldots 16}$ in the laboratory frame are obtained as a function of $H_0$ size and orientation.

\begin{figure*}[t]
	\centering
	\includegraphics[width=2\columnwidth]{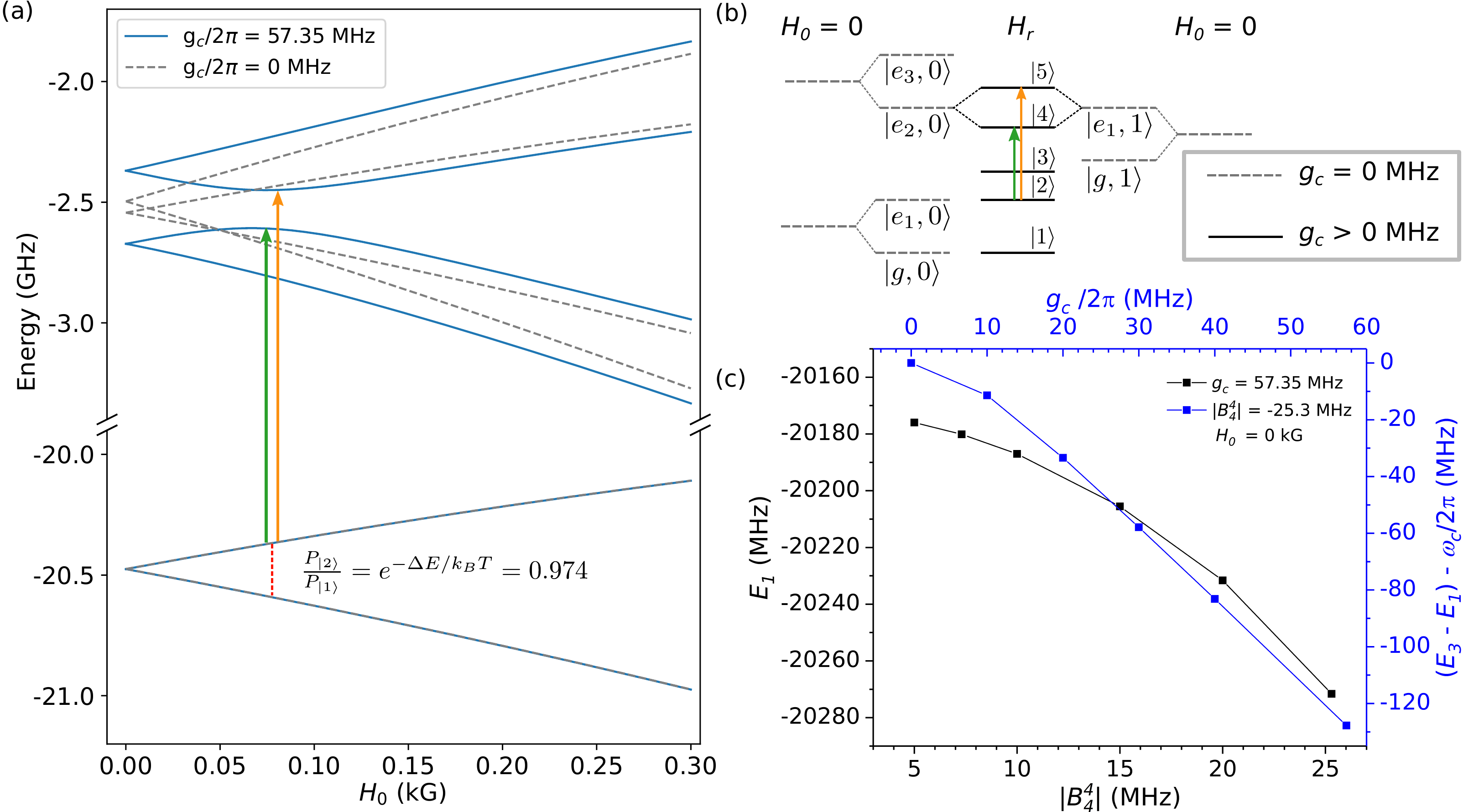}
	\caption{(a) The lowest six spin-resonator eigenenergies. For $g_c$~=~0 (dark gray) the states are pure spin states: ground $|g,0\rangle$ and excited $ |\left\{e_1,e_2,e_3\right\},0\rangle$ states for zero photons and $|g,1\rangle$, $|e_1,1\rangle$ for one photon in the resonator. For a large coupling $g_c/2\pi$~=~57.35~MHz (blue lines), a hybridization of $| e_2,0\rangle$ and $|e_1,1\rangle$ is indicated by the green and orange arrows. (b) Energy ladder diagram (not at scale). The $g_c=0$ states are shown by the dashed grey lines for the cases of 0 (left) and 1 (right) photons. The hybridized states ($g_c>0$) at resonance are represented by the solid black lines: $|2\rangle \rightarrow |4\rangle$ and $|2\rangle \rightarrow |5\rangle$ corresponds to the experimentally observed resonances. (c) Calculated perturbation of the spin-cavity ground state at zero field due to: (i) the crystal-field term $B_4^4$ with no photons in the cavity (left axis, black squares) and (ii) a single photon in the cavity seen as $(E_3-E_1)-\omega_c/(2\pi)$ as a function of $g_c/(2\pi)$ (right axis, blue dots); the two vertical axes have the same range. The two dependencies are similar, which indicates that the change in $B^4_4$ is due to the presence of a photon field in the cavity.} 
	\label{fig3}
\end{figure*}

In order to reproduce the observed signals we analyze the $|2\rangle\rightarrow|4\rangle$ and  $|2\rangle\rightarrow|5\rangle$ transitions, indicated in Fig. \ref{fig3} with green and orange arrows, respectively. We compute the absorption spectra, as the one shown in Fig~\ref{fig2}(b), using the transition amplitude: 
\begin{equation}
A_{if} = |\langle \psi_f|a|\psi_i\rangle|^2 e^{-E_i/(k_BT)}\delta\left(E_f-E_i - \hbar\omega\right)
\label{eq:Aif}
\end{equation}
for which a maximum value implies a photon absorption and thus a minimum in the $|S_{11}|^2$. The state $|\psi_i\rangle=|2\rangle$ is the first excited pure spin state; $|\psi_f\rangle=|4\rangle$ and $|\psi_f\rangle=|5\rangle$ correspond to the final states for the lower and upper branch transitions, respectively (Fig.~\ref{fig2}b). These states are shown in Table~\ref{table1} and obtained by exact diagonalization using the Quantum Toolbox in Python (QuTiP)~\cite{qutip2012}.

An iterative least-square comparison between the experimental spectra (Fig.~\ref{fig2}a) and calculated transition frequencies (Fig.~\ref{fig2}b) leads to crystal field parameters $B^q_k$, cavity resonance frequency $\omega_c/(2\pi)$, relative orientation $\theta$ of the static field relative to the \textit{c}-axis and spin-cavity coupling strength $g_c$: $B_2^0 =$~-945.66, $B_4^0=$~-1.2435, $B_4^4 =$~-25.3, $B_6^0 =$~5.712~$\times$10$^{-4}$ and $B_6^4 =$~70.0~$\times$10$^{-4}$ (all in MHz units), $\omega_c/(2\pi)=$~17930.7~MHz, $\theta=$~81.66$^{\circ}$ and $g_c=$~57.35~MHz. All crystal field parameters agree with our previous weak coupling study~\cite{Franco-Rivera2022} 
within 1\% difference except for $B^4_4$ which here is found to be $3.5\times$ larger. 

The six lowest eigenenergies $|k\rangle_{k=1\ldots6}$ are shown in Fig.~\ref{fig3}a for $g_c$ =~0 (dashed gray) and $g_c$ =~57.35~MHz (blue). For $g_c=$~0 the states are labeled $|g,n_c\rangle$ and $|e_j,n_c\rangle$ with $n_c=0,1$ for the ground and $j^{\mathrm{th}}$ excited spin state, respectively ($j=1,2,3$). States $|\left\{g,e_1\right\},0\rangle$ and $|\left\{e_2,e_3\right\},0\rangle$ correspond to the 7/2 and 5/2 Kramers doublets, respectively (see Fig. \ref{fig3}b left) split by the transverse field $H_0$. States $|\left\{g,e_1\right\},1\rangle$ correspond to the $|S_z|\cong7/2$ doublet raised by $\hbar\omega_c$ (see Fig. \ref{fig3}b right). A spin-cavity coupling strength $g_c/(2\pi)=$~57.35~MHz leaves $|1\rangle=|g,0\rangle$ and $|2\rangle=|e_1,0\rangle$ untouched but gives a hybridization between the $|e_1,1\rangle$ and $|e_2,0\rangle$ leading to the experimentally observed gap of 146~MHz between states $|4\rangle$ and $|5\rangle$ (states $|\{3,4,5\}\rangle$ are listed in Table~\ref{table1}). Thus, the exact diagonalization leads to a cooperativity factor $C=\frac{g_c^2}{\kappa_c\gamma_s}\approx35$ rather than $\approx58$ obtained from the two-level picture of the $S_{11}$ fit . We note that larger factors have been inferred in other RE ions~\cite{Wang2022} but no gap was observed.

The observed large difference in the crystal-field $B_4^4$ parameter between the weak and strong coupling cases, as mentioned above, can be explained using the Dicke Hamiltonian of Eq.~\ref{eq:gdcavH}. We calculate the change in $E_1\cong E_{g,0}$ at zero field as a function of $B_4^4$ (see Figure~\ref{fig3}c left axis) and compared it to the change inflicted by the presence of a single photon in the cavity $E_3-E_1$ as a function of $g_c$ for fixed $B_4^4=-25.3$~MHz ($E_3=E_{g,1}$ only for $g_c=0$). For a proper comparison, the photon frequency needs to be subtracted as well as shown in Fig.~\ref{fig3}c (right axis). The two vertical axes of Fig.~\ref{fig3}c have the same range: as $B_4^4$ or $g_c$ increases, so does the perturbation of the electronic ground state leading to a $\sim$130~MHz shift for the measured values of $B_4^4$ and $g_c$. The similarity between the two trends indicates that the photon field is so strongly coupled to spin that the electrostatic interaction, intrinsic to the crystal only, it's severely perturbed.

\begin{figure*}[t]
	\centering
	\includegraphics[width=2\columnwidth]{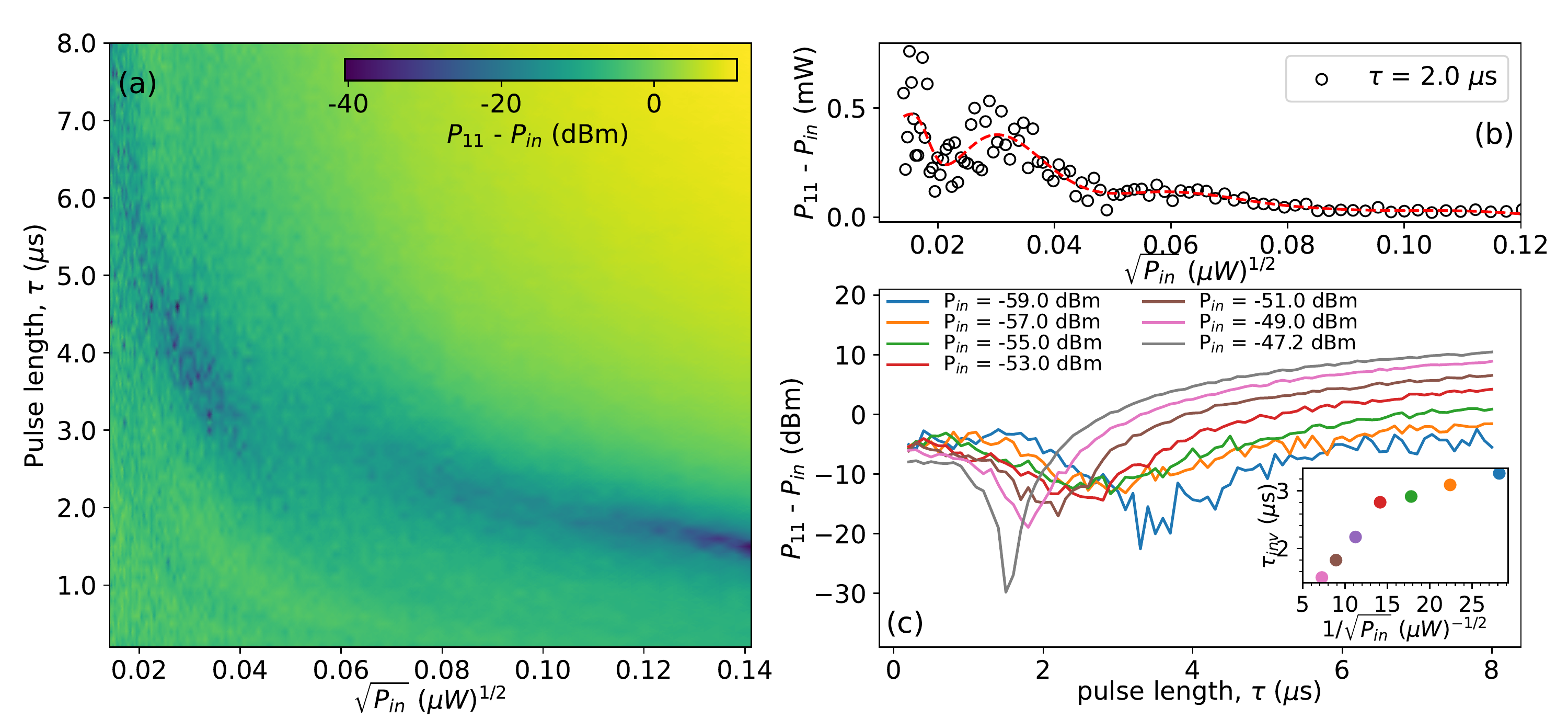}
	\caption{(a) Pulsed ESR measurements of the reflected cavity ringdown signal as a function of pulse length and squared-root of the applied drive power. (b) Horizontal cut of the FFT amplitude for a 2$\mu$s pulse and its low-pass FFT average done to remove unwanted noise: a clear pattern corresponding to the Rabi oscillations of the state admixture is seen. (c) Pulse length cuts of the relative FFT amplitude obtained from panel (a) at large applied powers $P_{in}$. Inset shows inversion pulse length $\tau_{\textnormal{inv}}$ vs the inverse of the cavity input drive power $1/\sqrt{P_{in}}$.} 
	\label{fig4}
\end{figure*}
\section{Single spin coupling strength estimation}
The average single spin coupling to the resonator is done by integrating the coupling to the vacuum field $\delta \boldsymbol{H}_1$ over the mode volume for an input power $P_{n=0}=\hbar \omega_c\kappa_c/2=$~-124.1~dBm. For $|\psi_{i,f}\rangle = |e_{1,2},0\rangle$ the single-spin coupling strength is $g_0(\mathbf{r}) = \gamma_e|\langle \psi_f | \delta\mathbf{H_1}(\mathbf{r})\cdot \mathbf{S} | \psi_i \rangle|$ where $\gamma_e=2.8025$~MHz/G and $\delta\boldsymbol{H}_1(\mathbf{r})$ is the spatial distribution of the vacuum fluctuations obtained from the COMSOL finite-element calculations as shown in Fig.\ref{fig1}(c)-(d). With a static field $\approx \perp c$-axis, the leading contribution to $g_0$ comes from $\langle e_2,0|S_{x}|e_1,0 \rangle=0.715$ while $\langle e_2,0|S_{z}|e_1,0 \rangle$ is only 0.01 (calculated with QuTiP) despite the fact that the $z$-field $\delta \text{H}_{1,z}$ contribution is $\approx$67\% larger than that for $\delta \text{H}_{1,x}$ (from COMSOL). The spin-ensemble coupling strength is~\cite{Franco-Rivera2022}:
\begin{equation}
	g_c = \gamma_e\left|\langle e_2,0|S_{x}|e_1,0 \rangle\right| \sqrt{\rho(0.7)\int_{V_m}^{}d^3\mathbf{r}\left|\delta \text{H}_{1,x}(\mathbf{r})\right|^2}
	\label{eq:gc}
\end{equation} 
where $V_m$ is the mode volume, the spin concentration $\rho$ is corrected by $min(\frac{\kappa_c}{\gamma_s},1)=1$ (spin selectivity as done by the cavity) and for the 70\% abundance of the Gd $I=0$ isotopes. The integration volume $V_m$ is increased around the center of the $\Omega$-loop and for $V_m\approx 10^4 \mu$m$^3$ ($\ll$ crystal volume) the experimental value of $g_c$ is obtained (see also \cite{Franco-Rivera2022} for a similar calculation); this volume corresponds to $N_s=7.5 \times 10^{9}$ excited spins. Therefore the spatially averaged single spin coupling strength is $g_0/(2\pi)=g_c/(2\pi\sqrt{N_s})\approx$ 620 Hz. An upper limit of the number of intra-resonator photons can be estimated from the input power at the sample holder level $P_{in}=-64$~dBm as $n_c=P_{in}/(\hbar\omega_c\kappa_c)=3.2\times 10^6$, thus fulfilling the limit $n_c \ll N_s$.

\section{Pulsed ESR with dressed states}
In our experiments, we excite state $|2\rangle$ rather than the ground state $|1\rangle$ due to the limited cavity bandwidth and matching between the resonance frequency and spin states. Both states are similar in the sense that they are spin states for $n_c=0$. We performed pulsed ESR measurements to study the dynamics of the $|2\rangle\rightarrow|4\rangle$ transition and explore a potential way of initializing the system in the $|1\rangle$ ground state. The spin-cavity system is initially in thermal equilibrium at $T$, with spins almost equally distributed between the $|1\rangle$ and $|2\rangle$ states ($\frac{p_{|2\rangle}}{p_{|1\rangle}}=e^{-\Delta E/(k_BT)}=0.97, \Delta E = 203$~MHz). A pulse $\omega/(2\pi)=17760$~MHz $\cong E_4-E_2$ (green arrow in Fig.~\ref{fig3}a) of variable duration $\tau$ and power $P_{in}$ is sent to the sample and the FFT of the reflected cavity ring down is analyzed~\cite{Schweiger2001,Chiorescu2010}. For each power value from $P_{in}=-67$~dBm to $-47$~dBm, $\tau$ is varied from 0.2 to 8$\mu$s. In Fig. \ref{fig4}a the reflected signal $P_{11}-P_{in}$ is plotted as a function of $\tau$ and $\sqrt{P_{in}}$. $P_{in}$ is subtracted from $P_{11}$ to show that the initial signal ($\tau \rightarrow 0$) is the same for all powers. A clear minima is observed (blue shade) while other oscillatory features are more vague but present nevertheless. 

This aspect is better shown in Fig.~\ref{fig4}(b) using the smoothing (dashed red line) of an horizontal cut at constant pulse length $\tau=$~2.0~$\mu$s which indicates a clear damped oscillation of the cavity-spin states. The main minimum is well developed at high powers, as shown by vertical cuts in Fig.~\ref{fig3}(c). The inset shows the position of the dip $\tau_{inv}$ as a function of $1/\sqrt{P_{in}}$ and gives the required pulse length to transfer the population of $|2\rangle$ to $|4\rangle$. In the simplest case of a driven two-level qubit, it is expected that the nutation angle increases linearly with $\sqrt{P_{in}}$ giving an oscillatory decay of the Rabi signal. In a similar fashion, the inversion time of the Rabi flop will decrease with increasing field amplitude. In the case of our experiments, the Gd spins are multi-level systems in the presence of a large and highly anisotropic crystal field (analyzing spin dynamics would require solving the time-dependence of the full Hamiltonian in the laboratory frame). An outcome of the observed population inversion suggests the possibility of an active cooling procedure \cite{Valenzuela2006, Leibfried2003}: (i) a $\tau_{inv}$ pulse depletes $|2\rangle$ and populates $|4\rangle$, (ii) a fast relaxation from $|4\rangle$ to $|1\rangle$ provides the cooling to the ground state. Such relaxation can be induced by a $\omega_{41}$ pulse resonant with the $|4\rangle\rightarrow|1\rangle$ transition or simply by waiting a certain amount of time. Although $\omega_{41}$ is slightly detuned from $\omega_c$, the pulse can have enough power to stimulate the $|4\rangle\rightarrow|1\rangle$ transition, by adding another microwave source. Using Eq.~\ref{eq:Aif} we find $A_{24}, A_{41}\gg A_{12}\cong0$ and thus the relaxation to $|1\rangle$ can take place prior to a repopulation of $|2\rangle$. By repeating the pulse sequence, one can in principle provide a spin initialization in its ground state.

The on-chip pulsed ESR measurements shown here represent a first step for the characterization of the spin dynamics of this multilevel system. Having a well known description of the pulse lengths and drive power will lead to $\pi$-pulse dynamical decoupling sequences such as spin echo\cite{Hahn1950} and Carr-Purcell-Meiboom-Gill (CPMG)\cite{Carr1954, Meiboom1958} that provide a direct measure of the spin dephasing time.
\section{Conclusion}
In conclusion, we demonstrate the strong coupling regime between the Gd$^{3+}$ multilevel spin system to an on-chip superconducting resonator. Using exact diagonalization of the spin-cavity Dicke Hamiltonian we estimate the coupling strength of the spin ensemble as $g_c/(2\pi)=$~57.35MHz. Together with the estimated cavity and spin dephasing rates, this leads to a cooperativity factor $C\approx$~35 and averaged single spin coupling strength of 620 Hz. The dynamics of the cavity-spin states are explored using pulsed ESR measurements and a population inversion is observed. Most remarkably, we measure a strong perturbation of the anisotropic crystal-field parameter $B_4^4$ due to the large coupling between the spins and the resonator.
\section*{Acknowledgements}
This work was performed at NHMFL at the Florida State University and supported by the National Science Foundation through Grant No. NSF/DMR-1644779 and the State of Florida. We acknowledge discussions with Dr. Petru Andrei (FSU) and we thank Dr. A.M. Tkachuk for providing the sample. S.B. acknowledges support from CNRS research infrastructure
INFRANALYTICS (FR2054) and the International Emerging Action QULT. S.M. acknowledges support from Grants-in-Aid for Scientific Research C (No. 18K03444) and the Elements Strategy Initiative Center for Magnetic Materials (ESICMM) funded by MEXT of Japan (Grant No. 12016013).

	%
	
	
\bibliography{onchip_Gd_strongcoupling.bib}
	
\end{document}